\documentclass[twocolumn]{article}
\begin{document}
\input{epsf}
 
\title{Temporal evolution of radiation from a plasma blob
 with constant deceleration: application to $\gamma$-ray bursts}
\author{Stanis{\l}aw Ry\'s and Krzysztof Ma\'slanka \\
      Astronomical Observatory of the Jagiellonian University \\
       $PL - 30 244$ KRAK\'OW  ul. Orla 171,  POLAND }
 
\date{12.08.2003}
\maketitle

\begin{abstract}
It has been widely accepted that relativistic effects have to be used
for modelling the wide classes of events associated with
active galactic nuclei and galactic superluminal sources.
 Here we propose a simple analytical model of decelerated motion which
is able to explain the shape of double-peaked light curves of
$\gamma $-ray bursts.
We show that the differences between individual bursts of this class 
may be due to the angle between the line-of-sight and 
direction of the outflow, and that the duration of the burst is 
intimately connected with the deceleration parameter.
Assuming special relativity formulae and the orientation of the observer
with respect to the outflow, we are able to model observed double-peaked
light curves and obtain good quality fits.
We propose that such smooth double peaked bursts could
be produced by new born galactic superluminal sources.
\end{abstract}


\section{Introduction}

 Only a small fraction of $\gamma $-bursts detected by BATSE (onboard
the Compton Gamma Ray Observatory) possess an optical identification.
 Therefore, most of the registered bursts do not have predictions for 
their total energy and we can expect that among unidentified bursts 
may exist some which happened close to us.
 Many active galactic nuclei (AGN) exhibit relativistic jets, and
the Galactic jet sources, or microquasars, also exhibit outflow in
the form of jets (Mirabel \& Rodriguez \cite{mir}).
  We can expect that lower energetic type of $\gamma$-ray bursts, 
may be connected with a class of new born galactic superluminal 
sources as it may by inferred from a recent review by Ghisellini and 
Celotti  (\cite{ghi} see also \cite{nor}).
  However, regardless of whether the radiation from a burst is isotropic or
anisotropic, an observer would see only a small fraction of the radiation
emitted in the burst, even if the observer is looking directly at the beam
which has finite angular width, he still sees only a fraction of the beam.
  The majority of radiation detected by the observer will be that having
the maximum relativistic intensity boosting in the direction of the
observer.
  Thus, the shape of the burst afterglow light curve should depend on
beaming effects (Rhoads \cite{rho}).

  The Lorentz time dilation between the frame of reference of the burst
and that of the observer, modifies the time-evolution of the observed
flux of photons.
It is known that radiating plasma eventualy decelerates (Piran \cite{pi1})
  Such a situation requires that we should derive a simple and fully
analytical model, which allows us to investigate the effects produced in
light curves by deceleration.
  We show below that the simplest model having constant deceleration can
explain particular light curves obtained with BATSE.
  The bursts detected by BATSE consist of either single, double or multiple
peaks in the light curve, but we should realize that such a
classification is determined by our resolution capabilities in
the time domain.
 Because we are interested in the influence of deceleration we will
avoid discussing any spatial structure of the source and we will assume 
that the temporal resolution of observational data is does not allow us
to detect spatial extension of the emitting region.
 In the following, we show that light curves exhibiting double peaks
can be explained in the framework of a simple model in a simple way.
  The double peaked light curve appears to be a generic property of
models with a decelerated motion.

\section{The model and its properties}
 
Consider the motion of a single blob of plasma in a relativistic jet
emerging from an active region (c.f. Dar \& R\'{u}jula \cite{dar}).
Suppose that this has the simplest form of decelerated motion, i.e.
motion with constant deceleration $g$ in a straight line - the simpliest 
assumption that could be made.
The agreement we find with observations (see Sec. 3), supports this 
assumption although the physical reason for constant deceleration is 
not clear.
One-dimensional analytical calculations give us 
(Rybicki \& Lightman \cite{rli}) the time dependence of the Lorentz 
factor $\gamma$ and velocity $\beta = v/c$ in the observer's frame:
 
\begin{equation}
\gamma =  \sqrt{1+ \left( \frac{\beta _0}{\sqrt{1-\beta _0^2 } }
            -\frac{gt}{c} \right)^{2}}
\label{jed}
\end{equation}
 
\begin{equation}
\beta  = \left( \frac{\beta _0}{\sqrt{1-\beta _{0}^{2}}}
        -\frac{gt}{c} \right) \cdot \frac{1}{\gamma (t)}
\label{dwa}
\end{equation}
 
\noindent where $\gamma _0 = \gamma (t=0)$, and $\beta _0 = \beta (t=0)$.
We also use the Doppler correction of light travel times in the form:
 
\begin{equation}
t - {r \over c} \cos \theta = T_{obs}
\label{trz}
\end{equation}
 
\noindent where $\theta $ is the angle between direction of motion and 
the line of sight, $t=(\gamma _0 \beta _0 -\gamma \beta )c/g$, and
$r=(\gamma _0 -\gamma)c^2 /g$, are quantities in the observer's frame,
and $T_{obs}$ is the time measured at the location of the observer.

The second assumption made is that the temporal evolution of the flux of
photons emitted during the burst is described by a power law
$F\sim \nu ^{-\alpha} T^{-\mu }$ in the rest frame of the blob.
Such a power law description of flux evolution is widely discussed and
frequently (assumed) obtained as an approximation of various emission 
mechanism in the literature (van Paradijs et al. \cite{pkw}).
Therefore the flux of photons received by the observer is:
 
\begin{equation}
F(t) = F_0 \cdot D^{3+\alpha } \cdot \tau ^{-\mu }
\label{czt}
\end{equation}
 
\noindent where the time elapsed in the frame of the emitter $\tau$ and
the Doppler boosting factor $D$ are given by:
 
\begin{equation}
\tau ={c\over g}\ln \left( \frac{(\beta _0 +1)\gamma _0}
  {(\beta +1) \gamma } \right)
\label{pie}
\end{equation}
 
\begin{equation}
D =\frac{1}{\gamma (t)\cdot (1-\beta (t)\cos \theta )}
\label{sze}
\end{equation}
 
The shape of the light curve implied by (\ref{czt}) is determined by
the values of $\cos \theta$, $\alpha $, $\mu $, and $\gamma _0$.
The deceleration parameter ($g/c$) has no influence on the curve shape
but only on the burst duration.
For any set of these parameters we can simply calculate that there is
a $T^{max}_{obs}$ after which the blob stopped its motion.
From (2), $\beta = 0$ implies that
$T_{obs}^{max} =[\gamma _0 \beta _0 -(\gamma _0 -1)\cos \theta]c/g$,
providing the value of $T_{obs}^{max}$.
When the blob has stopped ($T_{obs} > T^{max}_{obs}$) we adopt (\ref{czt})
to describe the burst simply by putting $D=1$.
At this point, times intervals measured in the source and observer frames
are equal.
Thus we are able to model light curves in terms of the parameters:
$\cos \theta $, $\alpha $, $\mu$ and $\gamma_0$.
 
 Figure 1 shows the properties of our model and the influence of various
parameters on the shape of the light curve, including the angle between
the velocity vector of the blob and the line of sight.
 In the case of small angles, we obtain a single burst with a `broken power
law' shape of light curve in the observer's frame, studied by several 
authors (e.g. Sari at al. \cite{spn}, Rhoads \cite{rho}, 
Panaitescu \& Kumar \cite{pak}).
 For larger angles we obtain no break in the light curve slope but simply
a brightening due to the debeaming effect (of the radiation) at later times
in the decay leading to the formation of a second observed peak.
 Thus, a double-peaked burst can be produced by a single blob.
The value of critical angle demarcating these effects depends on
the spectral index $\alpha $ of the burst and its decay index $\mu $.
 One can also see that a larger $\gamma _0$ leads a to a larger ratio
between the second peak intensity and the minimum intensity between
the two peaks.
 
\begin{figure} 
\epsfbox[25 65 250 680]{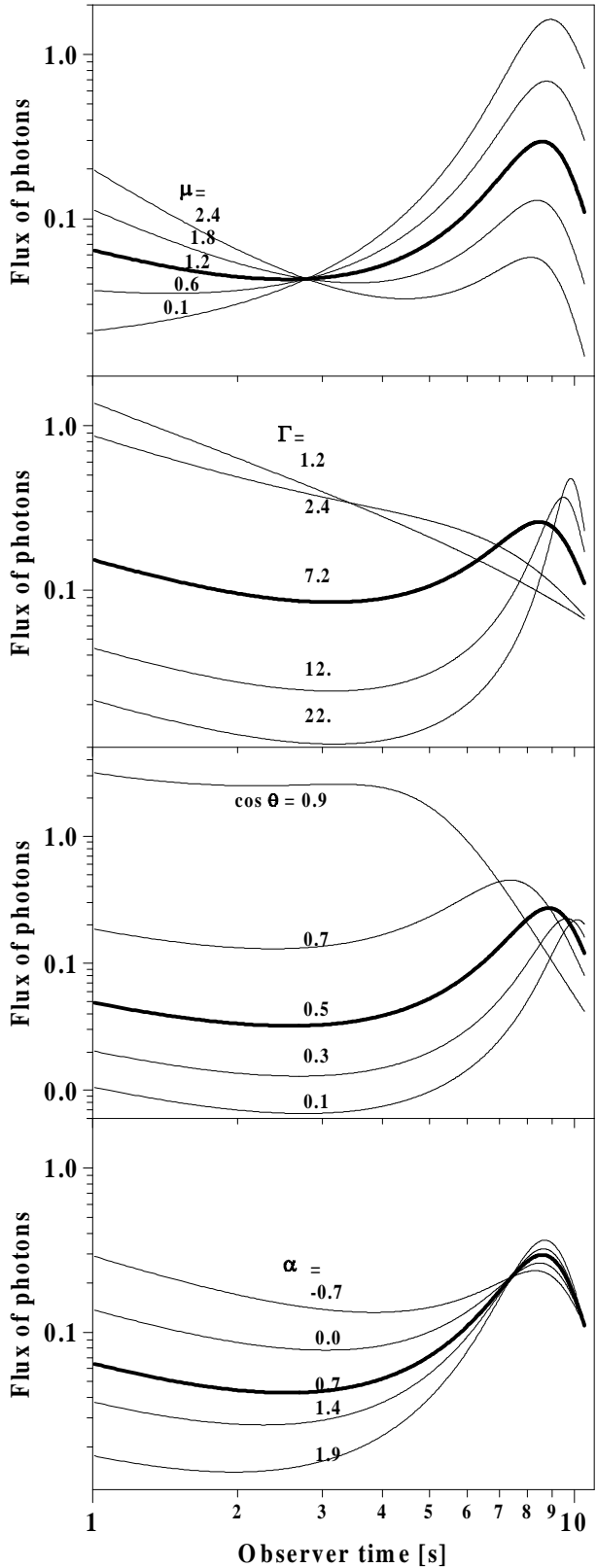}
\caption{Properties of the model i.e. the dependence of a light curve
shape on parameters $\mu $, $\Gamma $, $\cos \theta $, and $\alpha $.
Heavy curves are those with
$\mu =1.2$, $\alpha =0.7$, $\gamma _0 =7.2$, $\cos \theta = 0.5$.
The curves end at $T=T^{max}_{obs}$ i.e. at the time when the blob
stopped its motion.
For different values of $\gamma _0$ and $\cos \theta $ we change the 
value of $g/c$ in order to make the same value of $T_{obs}^{max}$ for
each curve.}
\end{figure}
 
\section{Two examples of fitting}
 
To test the model, we selected two double-peaked $\gamma $-ray light
curves from the BATSE archive (Mallozii \& Six \cite{mal}) which are 
thus consistent with being produced by a single blob.
 We chose the bursts which took place on 1999, May 31 and 2000, May 19
designated GRB$\, 990531$ and GRB$\, 000519$.
 We fitted the data in each case with the above model defined by
(\ref{czt}) using a fitting procedure in which each of the 4 free
parameters is assigned random values by a Monte Carlo procedure,
this being repeated many times until an acceptable fit is obtained
(i.e. $RMS_{fit} < 3\times RMS$ of the fluctuation before the burst).
Because we adopted the simplest model of evolution with a singularity
at $\tau = 0$, (i.e. $T_{obs}=0$, - the trigger time) we were unable
to calculate the growth phase of the burst and we fitted only its
descending phase from the first peak (from $T_{obs}=T_0 >0$) and
the entire second peak.
The results of fitting are shown in Figure 2 \& 3 and in Table 1.
 
\begin{table}
\caption{Fitted values of parameters. In paretheses are values obtained
by independent fitting of the model with the asumption
$\gamma_{end}=10 $ (see text).}
\begin{center}
\begin{tabular}{lrlrl}
\hline\noalign{\smallskip}
GRB                & 990531 &        & 000519 &  \\
\noalign{\smallskip\hrule\smallskip}
BATSE trigger      & 7975   &        & 8111  &       \\
$\gamma _0$        & 15.2   & (119.) & 7.0   &(35.8)  \\
$\cos \theta $     & 0.183  & (.985) & 0.070 &(.847)  \\
$\alpha $          & 1.42   & (1.97) & -0.15 &(1.36)  \\
$\mu $             & 3.11   & (3.18) & 1.46  &(1.65)  \\
g/c                & 0.489  & (.067) & 1.113 &(.747)  \\
$T_0$              & 2.071  & (2.02) & 0.679 &(.611)  \\
\noalign{\smallskip}
\hline
\end{tabular}
\end{center}
\end{table}
 
In both events, the shape of the light curve near the second peak
(see Figs. 2 \& 3) is determined physically by the debeaming process
(as explained above in the discussion of Fig.1)
when most of the bulk kinetic energy was transformed into thermal energy
of the surrounding matter via shock interactions, according to the model
of M\'esz\'aros \& Rees (\cite{mr2}, see also Piran \cite{pi1}).
 We suggest that this is the time when optical and/or radio emission could
rise to a detectable level.
Because in the case of strong deceleration, the second peak is located
a few seconds after the first peak, we are unable to see their afterglow
as optical/radio transients since there is not enough time after the high
energy detection to move the telescopes to the correct coordinates
(Paczy\'nski \& Rhoads \cite{par}).
 
\section{Discussion}
Our model describes a point-like element of plasma moving with relativistic
speed.
 Using a Lagrangian description of fluid motion we could construct a model
of more complicated flow similar to a supernova shell during its explosion
(Colgate \cite{col}, Galama et al. \cite{gal}).
 In this case we would observe different parts of the shell possessing
different values of the parameter cos($\theta$) and the observed light
curve shape would be a superposition of the single burst light curve models.
 Since the light curve data are fitted well using a model involving only
a single blob of plasma, we do not have to assume that the shape of
the moving volume influences the observed light curve as proposed by
Moderski et al. (\cite{msb}).
 
 We realize that the low value of the Lorentz factors derived in
this paper are inconsistent with all but the very softest gamma-ray spectra
(Woods \& Loeb \cite{woo}, Lithwick \& Sari \cite{lit}).
However, our model may be interpreted from the viewpoint of an internal
shock model scenario (Kobayashi at al. \cite{kob}, Piran \cite{pi1}, 
Dar \& R\'{u}jula \cite{dar}) which assumes that moving blobs (or shells) 
possess different velocities which lead to collisions between them.
 Each collision generates a relativistic shock which radiates and
decelerates continually.
 In order to compare this scenario with our model we need to introduce
a new parameter $\gamma_{end}$ (which is the Lorentz factor after the 
deceleration is switched off), but we are unable to determine its value
because the corrections produced by $\gamma_{end}$ take the form of
a scaling factor (relative to $g$).
 This means that we are able to determine it only when we have some
predictions of total (peak) power or the true time scale of the burst
phenomenon.
In the case of a burst having an afterglow, it is obvious that
$\gamma_{end}$ has a value much greater than 1.
Such a value may be calculated as the initial value in the model of
a particular afterglow.
 Therefore values given in Table 1 should be scaled in the case of
comparison with true physical conditions inside the shock and $\gamma_0$
represents the difference in Lorentz factors between two collided blobs.
 In particular, this should be done in the majority of GRBs which could not
be explained by a low value of $\gamma_0$.
  As an example, in Table 1 we put in parentheses the values obtained
during independent fitting of the model in the case of $\gamma_{end}=10$.
\begin{figure} 
\epsfbox[85 65 300 260]{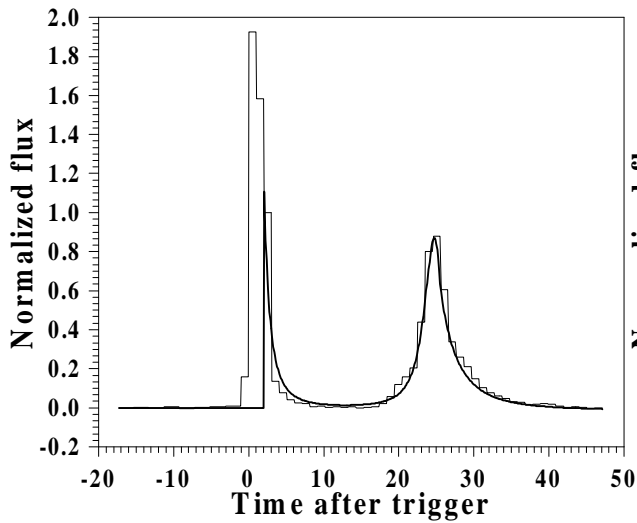}
\caption{The observed and the best fitted light curves for $\gamma$-ray
burst T7592. The model has parameters: $\gamma _0=15.2$, $\mu =3.11$, 
$\alpha =1.42$, $\cos (\theta )=.183$.}
\end{figure}
\begin{figure} 
\epsfbox[85 65 300 260]{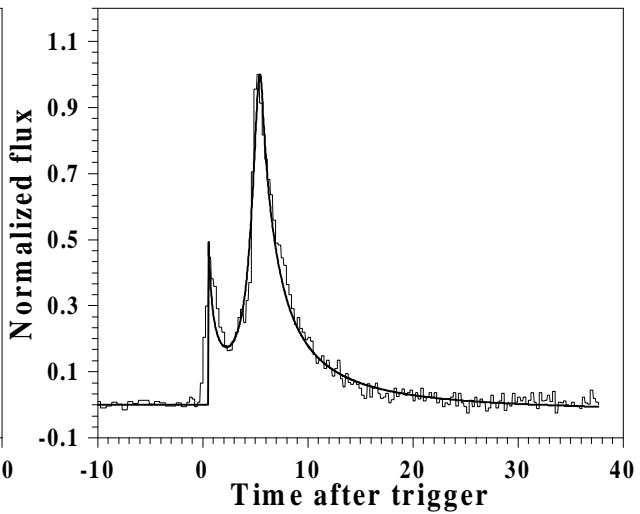}
\caption{The observed and the best fitted model light curves in 
$\gamma$-ray burst T8111. The model has parameters: $\gamma _0=7.0$, 
$\mu =1.46$, $\alpha =-0.15$, $\cos (\theta )=.070$.}
\end{figure}
 
 Given that GRBs will have a distribution in space, it is clear that some
bursts will take place relatively close to us.
 If we accept that the bursters produce an extreme relativistic outflow,
it is more likely that the outflows will move in a direction different
from the line of sight to the observer, compared with a more distant burst
which we only see if there is alignement (Paczy\'nski \cite{pac1}).
 In such a case,  we will see the burst light curves that are strongly
modified by relativistic effects, but with small boosting in the luminosity.
 Because we obtain small values of $\cos \theta $ and $\gamma _0$ by
fitting, we deduce that the two examples of light curves presented in
Figures 2 \& 3 belong to this type of burst and they are much closer to us
than bursts for which redshifts have been determined 
(M\'esz\'aros \cite{mes}).

 Since we obtain low values of $\cos \theta $, therefore the outflow 
direction is far from the line of sight and we have no Doppler boosting 
of radiation in the observer's frame.
 Energy considerations lead us to the conclusion that the examples of 
light curves discused here, were produced by objects close to us than 
those detected at cosmological distance.
 If we have a burst event at a distance of a few $kpc$ in comparison to 
$Gpc$, the flux can be $(10^{12}$ times larger which allows us to
see an event which is not boosted but dimmed by relativistic effects 
(or may be also less energetic).
 A scaling factor of $10^{12}$ allows us to calculate the upper
limit for the Lorentz factor as: 
$10^{12} > (\gamma ^4)^2 $ ; i.e. $\gamma < 10^{1,5} \sim 30$.
 We use $\gamma ^4$ twice because firstly we have no Doppler bosting
and secondly because we have Doppler dimming.

 It is well known that galactic microquasars (like GRS1915+105) 
eject blobs with relativistic speed (Mirabel \& Rodriguez \cite{mir}).
 If we assume that similar sources are born in a much more dense environment
and work in a similar manner ejecting blobs which are abruptly stopped
leading to an explosive transfer of kinetic energy into radiation, 
we have a physical interpretation of the results obtained.

\section{Summary}
 This paper discusses the simplest model of the influence of 
deceleration on the observed shape of the light curves of $\gamma $-ray
bursts.
 Taking into account the deceleration, we obtain important modification
of the power law decaying light curve. 
The transformation of times between the proper frame of a blob and 
the observer's frame, i.e. a logarithmic dependence (Eq. 5), may be
interpreted as an amplification of the time interval, i.e. at the begining
of the burst we can observe processes lasting few milliseconds as events
lasting a few seconds.

In any case of emission, a spatialy extended region could be modelled as
a superposition of point-like sources. 
Therefore the shape of the light curve of a point-like source give us 
the limit for the fastest temporal variations of the flux.
This fact will lead us to the prediction that all spatially extended flows 
should possess more smooth light curves than predicted by our model.
 In the framework of a decelerated motion model, we are able to produce 
shorter bursts (i.e. with lower value of $T^{max}_{obs}$) if we assume
stronger deceleration.
All events shorter than the pulse of radiation discussed here, i.e. induced
by a deceleration process, probably come from single events of coherent 
emission, but for discussion of such an event we should first separate
the influence of special relativity effects from observational data.
Therefore when one postulates a spatial shape of the outflow 
(e.g. jet type), the temporal properties of the emiting region 
should be discussed first with inclusion that the decelerated motion 
gives us the possibility of seeing the millisecond events expanded to 
more than a few seconds.

 Values of $\gamma _0$ obtained by fitting suggest that the events under 
consideration belong to a lower energetic type of $\gamma$-ray burst, 
which may be connected with a class of new born galactic superluminal 
sources (c.f. \cite{ghi, nor}).
 


\begin{thebibliography}{}
\bibitem{col}
  Colgate, S.A. 1974, ApJ, 187, 333
\bibitem{dar}
  Dar, A., de R\'{u}jula, A. 2000, ($/astro-ph/0008474$
)
\bibitem{gal}
  Galama, T.J. et al. 1998, Nature, 395, 670
\bibitem{ghi}
  Ghisellini, G. \& Celotti A. 2002, in: Blazar astrophysics with BeppoSAX
  and other observatories, workshop ($/astro-ph/0204333$) 
\bibitem{kob}
 Kobayashi, S., Piran, T., Sari, R. 1997, ApJ, 490, L92
\bibitem{lit}
  Lithwick, Y.  \& Sari, R. 2001, ApJ, 555, 540
\bibitem{mal}
  Mallozzi, R.S.  \& Six, F. $http://gammaray.msfc.nasa.gov/$
\bibitem{mes}
  M\'esz\'aros, P. 2001, Science, 291, 79
\bibitem{mr2}
  M\'esz\'aros, P.  \& Rees, M.J. 1999, MNRAS, 306, L39
\bibitem{mir}
  Mirabel, F. \& Rodriguez, L.,1999, ARA\&A, 37, 409 
\bibitem{msb}
  Moderski, R., Sikora, M.  \& Bulik, T. 2000, ApJ, 529, 151
\bibitem{nor}
  Norris, J.P., 2003, in: The Astrophysics of Gravitational Wave 
  Sources, workshop ($/astro-ph/0307279$)
\bibitem{pac1}
  Paczy\'nski, B., 2001, Acta Astron., 51, 1
\bibitem{par}
  Paczy\'nski, B.  \& Rhoads, J.E. 1993, ApJ, 418, L5
\bibitem{pak}
  Panaitescu, A.  \& Kumar, P. 2000, ApJ, 543, 66
\bibitem{pi1}
  Piran, T. 1999, Phys.Rep., 314, 575
\bibitem{rho}
  Rhoads, J.E. 1999, ApJ, 525, 737
\bibitem{rli}
  Rybicki, G.B.  \& Lightman, A.P. 1979, Radiative Processes 
  in Astrophysics (John Wiley \& Sons Inc.)
\bibitem{sph}
  Sari, R., Piran, T.  \& Halpern, J.P. 1999, ApJ, 519, L17
\bibitem{spn}
  Sari, R., Piran, T.  \& Narayan, R. 1998, ApJ, 497, L17
\bibitem{pkw}
  van Paradijs, J., Kouveliotou, Ch.  \& Wijers, R.A.M.J. 2000,
  ARA\&A, 38, 379
\bibitem{woo}
  Woods, E.  \& Loeb, A. 1995, ApJ, 453, 583
\end{thebibliography}
\end{document}